\documentclass[prd,aps,nofootinbib,preprint,showpacs,  tightenlines]{revtex4}

\usepackage{graphicx}

\begin{document}

\title{Axion inflation with gauge field production and \\ primordial black holes}

\author{Edgar Bugaev}
\email[e-mail: ]{bugaev@pcbai10.inr.ruhep.ru}

\author{Peter Klimai}
\email[e-mail: ]{pklimai@gmail.com}
\affiliation{Institute for Nuclear Research, Russian Academy of
Sciences, 60th October Anniversary Prospect 7a, 117312 Moscow, Russia}




\begin{abstract}
We study the process of primordial black hole (PBH) formation
at the beginning of radiation era for the cosmological
scenario in which the inflaton is a pseudo-Nambu-Goldstone
boson (axion) and there is a coupling of the inflaton
with some gauge field. In this model inflation is
accompanied by the gauge quanta production and a strong
rise of the curvature power spectrum amplitude at small
scales (along with non-Gaussianity) is predicted. We
show that data on PBH searches can be used for a derivation
of essential constraints on the model parameters in
such an axion inflation scenario. We compare our numerical
results with the similar results published earlier,
in the work by Linde et al.
\end{abstract}



\maketitle

\section{Introduction}
\label{sec-intro}

It is well known that inflationary models which predict prolonged inflation
are very  sensitive to Planck-scale physics (see, e.g., recent reviews
\cite{Baumann:2009ds, Pajer:2013fsa}). This sensitivity is especially important
for large field models when one needs to protect the inflationary potential
from a possible large effect of an infinite number of higher dimension
operators. Even in supersymmetric models of inflation this protection
is not guaranteed, because the supersymmetry is broken by the inflationary
background at the Hubble scale.

It had been shown very long ago that the simplest and most natural solution of
this problem is to assume that the inflaton $\varphi$ is a pseudo-Nambu-Goldstone
boson (PNGB) \cite{Freese:1990rb, Adams:1992bn, Kim:2004rp, Dimopoulos:2005ac,
Easther:2005zr, McAllister:2008hb, Flauger:2009ab, Anber:2009ua, Kaloper:2008fb,
Kaloper:2011jz, Berg:2009tg}, because in this case there is the shift symmetry,
$\varphi\to \varphi + {\rm const}$, broken by instanton effects (or explicitly).
In the limit when this symmetry is exact, the potential is flat, and the
corrections to slow-roll parameters are under control due to the smallness
of the symmetry breaking.

If PNGB is pseudoscalar (e.g., it is axion), it is natural to assume that there is a coupling
of it with some gauge field. This coupling is not forbidden by the shift
symmetry and, in general, is phenomenologically favourable (e.g., it can
lead to successful reheating). This coupling is essential if the axion decay
constant, $f$, is not too large (because the interaction term is inversely
proportional to $f$, see Eq. (\ref{Lint}) below). In UV-complete models
of axion inflation (e.g., those based on the string
theory \cite{McAllister:2008hb}) one has $f \ll M_P$ and, at the same time,
large excursion of the axion field is allowed. The inflationary potential
in these models is similar with the potential in large field models.

The main feature of the axion inflation with inflaton-gauge field coupling
is that such a coupling leads to a production of gauge quanta, and, through
the inverse decay of these quanta into inflaton perturbations, to a rise
of non-Gaussianity effects\footnote{Non-Gaussian effects in processes of
PBH formation had been studied in several pioneering works
\cite{Bullock:1996at, Ivanov:1997ia, Yokoyama:1998pt, PinaAvelino:2005rm, Saito:2008em}.}
and violation of scale-invariance.
In particular, a rather essential formation of primordial black holes (PBHs)
becomes possible \cite{Lin:2012gs, Linde:2012bt}
\footnote{Inflation models with PNG-fields coexisting with inflaton,
and subsequent PBH production processes, had been considered in
\cite{Rubin:2001yw, Khlopov:2004sc}.}.

In the present work we consider a process of PBH formation and PBH constraints
for the axion inflation models in which the inflationary expansion is
accompanied by the gauge quanta production. Our consideration differs
from the consideration carried out in the recent work \cite{Linde:2012bt}
in two respects. Firstly, we checked the hypothesis that a probability
distribution function (PDF) for curvature fluctuations produced in
axion inflation model has the same form as in $\chi_n^2$-models.
Secondly, for calculation of the $\beta_{PBH}$-functions describing the
fraction of the Universe's mass in PBHs, at their formation time, we
use the full machinery of the Press-Schechter \cite{PS} formalism
rather than the simple integral over the PDF of the curvature field
(see, in this connection, works \cite{Bugaev:2008gw, Bugaev:2010bb}).

The plan of the paper is as follows. In the second section we review
the main assumptions and formulas of the axion inflation model in
which there is a coupling of the inflaton with the gauge field.
In the third section, we discuss the choice of a suitable PDF for the
$\zeta$-field in our scenario. In the fourth section
we, using the Press-Schechter formalism, derive PBH mass spectra
needed for an obtaining the PBH constraints. Last section contains our
conclusions. In Appendix \ref{sec-app-hor} we study a time evolution of the curvature
perturbation power spectrum behind the Hubble horizon.
In Appendix \ref{sec-app-bisp} we study the shape of the $\zeta$-bispectrum
in our axion inflation scenario, comparing it with the prediction
of $\chi^2$-model.

\section{Axion inflation with gauge field production}
\label{sec-2}

\subsection{Outline of the model}

We consider the model of axion inflation in which there is a coupling
of the pseudoscalar inflaton (axion) to gauge fields of the form
\begin{equation}
\label{Lint}
{\cal L}_{int} = - \frac{\alpha}{4 f} \varphi F_{\mu\nu} \tilde F^{\mu\nu} \; ,
\end{equation}
where $F_{\mu\nu}=\partial_\mu A_\nu - \partial_\nu A_\mu$
is the field strength corresponding to some $U(1)$ gauge field $A_\mu$,
and
$\tilde F^{\mu\nu}=\eta^{\mu\nu\omega\theta} F_{\omega\theta}/(2 \sqrt{-g})$
is the dual strength, $f$ is the axion decay constant, $\alpha$ is the
dimensionless parameter.

It had been shown in \cite{Anber:2009ua} that the evolution (rolling) of
the inflaton leads to a generation of the field $A_\mu$ and to a subsequent
amplification (due to tachyonic instability) of its modes.
The solutions for the amplified modes are well parameterized by
the formula (index $+$ means the circular polarization of quanta)
\begin{equation}
\label{tildeAplus}
\tilde A_{+} (k, \tau) \cong \frac{1}{\sqrt{2k}} \left( \frac{k}{2 \xi a H} \right)^{1/4}
\exp\left[ \pi\xi - 2 \sqrt{\frac{2 \xi k}{a H} } \right]\; ,
\end{equation}
where
\begin{equation} \label{xi-equiv}
\xi \equiv \frac{\alpha \dot\varphi} {2 f H}\; ,
\end{equation}
and $\tau \cong -1/(aH)$. During inflationary expansion the value of $\xi$
changes with time. If $\xi$ is larger than $1$, the amplification factor $e^{\pi\xi}$
is essential. The production of gauge field quanta can affect the inflationary
process. In general, it prolongs inflation \cite{Anber:2009ua} because
it sources inflaton perturbations through the inverse decay:
$\delta A + \delta A \to \delta \varphi$ \cite{Barnaby:2010vf}.

The tachyonic amplification of gauge field modes leads to a characteristic
evolution of the power spectrum of primordial curvature perturbations.
The production of gauge quanta causes strong increasing of the
spectrum amplitude. To put constraints on this increase from PBHs one must
study the behavior of $\xi$-parameter as a function of time during inflationary
expansion. The cosmological evolution equations for the inflaton with extra
contributions from the gauge field are \cite{Anber:2009ua}
\begin{equation} \label{ddotphi}
\ddot\varphi + 3 H \dot \varphi + V' = \frac{\alpha}{f} \langle \vec E \cdot \vec B \rangle \; ,
\end{equation}
 \begin{equation} \label{fr}
3 H^2 M_P^2 = \frac{1}{2} \dot \varphi^2 + V + \frac{1}{2} \langle \vec E^2 + \vec B^2 \rangle.
\end{equation}
Here,
\begin{equation}
\vec B \equiv \frac{1}{a^2} \vec \nabla \times \vec A \; , \qquad \vec E \equiv - \frac{1}{a^2} \vec A' \;.
\end{equation}
The connection of $\langle \vec E \cdot \vec B \rangle$ and $\langle \vec E^2 + \vec B^2 \rangle$
with $\xi$ is given by \cite{Anber:2009ua}
\begin{equation}
\langle \vec E \cdot \vec B \rangle \approx - 2.4 \times 10^{-4} \; \frac{H^4}{\xi^4} e^{2\pi \xi}\;, \qquad
\frac{1}{2} \langle \vec E^2 + \vec B^2 \rangle \approx 1.4  \times 10^{-4} \; \frac{H^4}{\xi^3} e^{2\pi \xi}\; .
\end{equation}

For a calculation of the curvature power spectrum one needs the evolution equation
for inflaton field perturbation, $\delta\varphi$. Deriving this equation one must
take into account the backreaction effects \cite{Anber:2009ua, Barnaby:2011qe}.
The approximate accounting of these effects leads to the (operator) equation
\cite{Anber:2009ua, Barnaby:2011qe}
\begin{equation}
\label{ddotbeta}
\delta \ddot \varphi + 3 \beta H \delta \dot \varphi -
\frac{\nabla^2}{a^2}\delta\varphi + V'' \delta \varphi =
\frac{\alpha}{f} \left[\vec E \cdot \vec B - \langle \vec E \cdot \vec B \rangle \right],
\end{equation}
where $\beta$ is defined by the expression
\begin{equation}
\label{beta-equiv}
\beta \equiv 1 - 2 \pi \xi \frac{\alpha}{f} \frac{\langle \vec E \cdot \vec B \rangle}{3 H \dot \varphi}.
\end{equation}

Equations (\ref{ddotphi}) and (\ref{fr}) are solved numerically, giving the solutions
$\varphi(t)$ and $H(t)$ with initial conditions for $\varphi(0)$ and $H(0)$,
where $t=0$ corresponds, in our case, to the moment when CMB scales exit the horizon.
As a byproduct one obtains the function $\xi(t)$.

\subsection{Axion potential}

A typical axion inflationary potential which is exploited in natural inflation
models \cite{Freese:1990rb, Adams:1992bn} is given by the formula
\begin{equation} \label{Vphi4}
V(\varphi) = \Lambda^4 \left[ 1- \cos\left( \frac{\varphi}{f} \right) \right].
\end{equation}
In UV-complete models of axion inflation, the axion action is shift-symmetric, i.e.,
the shift symmetry $\varphi \to \varphi + {\rm const}$ is broken only non-perturbatively. In
particular, in closed string models with type \rm{II}B Calabi-Yau orientifold
compactifications such axions are available (see, e.g., the review paper
\cite{Cicoli:2011zz}). The inflationary potential in such models is periodic,
due to instanton effects, but it is flat enough for driving inflation only
in the case when the axion decay constant is larger than $M_P$. It is well known, however,
that it is difficult to obtain such large values of $f$ in UV-complete
theories \cite{Banks:2003sx, ArkaniHamed:2006dz}. So, the potential of
single axion, Eq. (\ref{Vphi4}), cannot provide the large field inflation
with long slow-roll evolution and a large value of the field excursion.

There are several groups of models in which the large field inflation
is possible with sub-Planckian axion decay constants: "Racetrack inflation" models
\cite{BlancoPillado:2004ns}, $N$-flation models \cite{Dimopoulos:2005ac},
assisted inflation models \cite{Liddle:1998jc, Copeland:1999cs}, axion monodromy
inflation models \cite{McAllister:2008hb, Flauger:2009ab, Berg:2009tg,
Kaloper:2008fb, Kaloper:2011jz}. The latter approach looks very promising
and we used it in the present paper for numerical calculations. In
particular, it had been shown in \cite{McAllister:2008hb} that, in
\rm{II}B string theory, the presence of space-filling $D_p$-branes
wrapping some two-cycles of the compact internal space leads to a breaking
of the shift symmetry and to the monodromy phenomenon: the potential
energy for the axion arising from integrating two-form fields over these
two-cycles is not periodic and increases with an increase of the axion field.
As a result, one has the additional component of the axion potential,
\begin{equation}
V(\varphi) = V_{sr}(\varphi) + V_{inst}(\varphi).
\end{equation}
Here, the abbreviation "sr" means slow-roll, and "inst" means instanton. In the concrete model
\cite{McAllister:2008hb}, with the $C_2$-axion and $NS5$-brane wrapping
$\Sigma_2$ (see \cite{Cicoli:2011zz} for notations), the potential $V_{sr}$
is given by the expression
\begin{equation}
V_{sr}(\varphi) = \frac{\epsilon}{g_s^2 (2\pi)^5 \alpha'^2}
 \sqrt{L^4 + g_s^2 \frac{\varphi^2}{f^2} }.
\end{equation}
Here, $L$ is the dimensionless modulus ($L^2$ is the size of the 2-cycle $\Sigma_2$),
$g_s$ is the string coupling constant, $1/(2\pi\alpha')$ is the string tension,
$\epsilon$ is the warp-factor \cite{McAllister:2008hb}. At large values of
$\varphi/f$ one has the linear potential,
\begin{equation} \label{V-linear}
V_{sr}(\varphi) \approx \mu^3 \varphi.
\end{equation}

The different realization of the monodromy idea (which is not based on the string
theory) had been suggested in \cite{Kaloper:2008fb, Kaloper:2011jz}. In these works,
the axion potential is generated by modification of the action introducing
there the coupling of the axion to a $4$-form. This new coupling leads to a spontaneous
breaking of the shift symmetry and to appearing (in the simplest case) the quadratic
axion potential just like in the original chaotic inflation scenario \cite{Linde:1983gd}.

In this work we will consider both cases: the axion inflation with the quadratic potential
\begin{equation} \label{V-square}
V(\varphi) = \frac{m^2\varphi^2}{2}
\end{equation}
(PBH constraints for axion inflationary model with such a potential have been
considered in work \cite{Linde:2012bt}) and
the inflation with the linear potential given by Eq. (\ref{V-linear}).
We assume that effects from the presence of $V_{inst}$ are subdominant and neglect
this term.

Using the expressions for axion potentials, Eqs. (\ref{ddotphi}) and
(\ref{fr}) can be solved.
The initial conditions for $t=0$ corresponding to the moment of time
when the scale with the comoving wave number $k=k_*=0.002\;{\rm Mpc}^{-1}$ enters
horizon are
\begin{equation}
\varphi(t=0)=\varphi_0 , \qquad \dot\varphi(t=0)=-\frac{V'(\varphi_0)}{3 H_0 },
\qquad H(t=0)  = H_0 = \frac{1}{M_P\sqrt{3} } V(\varphi_0)^{1/2}.
\end{equation}
The constant $m$ (or $\mu$) is fixed by the requirement that the curvature perturbation
power spectrum ${\cal P}_\zeta$ reaches the observed value \cite{Ade:2013uln} at
cosmological scales, ${\cal P}_\zeta(k_*)\approx 2.4\times 10^{-9}$.
For the linear potential (\ref{V-linear}), we obtained $\mu\approx 6.3\times 10^{-4} M_P$
and the following set of initial conditions: $\varphi_0\approx 10.6 M_P$,
$|\dot\varphi_0|\approx 2.8\times 10^{-6} M_P^2$, $H_0\approx 2.9\times 10^{-5} M_P$.
For quadratic  potential (\ref{V-square}), we have $m=6.8\times 10^{-6} M_P$ and $\varphi_0\approx 15 M_P$,
$|\dot\varphi_0|\approx 5.6\times 10^{-6} M_P^2$, $H_0\approx 4.2\times 10^{-5} M_P$.

\begin{figure}[!t]
\center %
\includegraphics[width=9 cm, trim = 0 0 0 0 ]{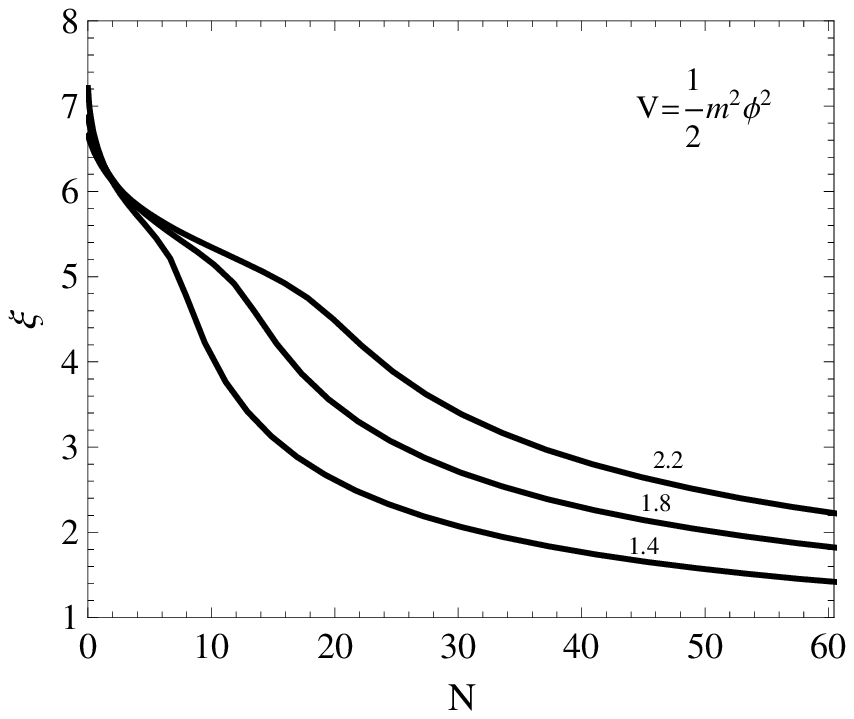}
\includegraphics[width=9 cm, trim = 0 0 0 -10 ]{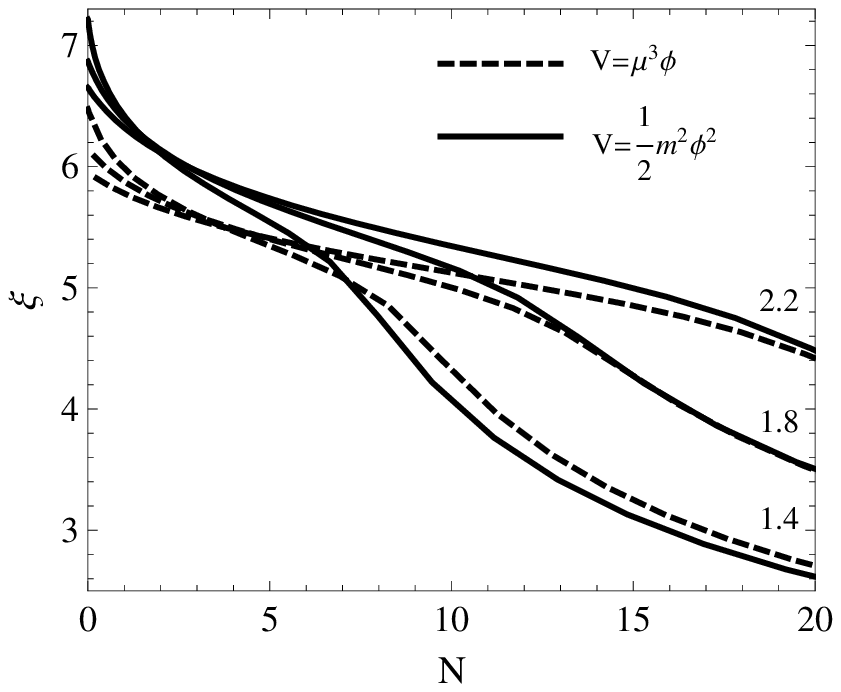}
\caption{ \label{fig-xi}
The value of $\xi(N)$ for different values of $\xi$ at CMB scales
and different choice of model potential. Solid curves are
for the case of potential (\ref{V-square}).
Dashed curves are for potential (\ref{V-linear}).
The curves are labeled with the value of $\xi(N_{CMB})$.
}
\end{figure}

In Fig. \ref{fig-xi} we show the results of our numerical
calculations: the dependence of $\xi$ on $N$, the number of e-folds before
an end of inflation, for different values of $\xi$ at CMB scales.

One should note, closing this subsection, that axion monodromy inflation with potentials given
by Eqs. (\ref{V-linear}) and (\ref{V-square}) predict rather large values of tensor-to-scalar
ratio: $r=0.07$ for the linear potential, and $r=0.14$ for the quadratic one. The latter
value is not excluded by the Planck \cite{Ade:2013uln} and BICEP2 \cite{Ade:2014xna} data.

\subsection{Curvature power spectrum}

In the limit of very small backreaction one has $\beta \to 1$. In this limit, the solution
of Eq. (\ref{ddotbeta}) is \cite{Barnaby:2010vf, Barnaby:2011vw}
\begin{equation}
{\cal P}_{\zeta}(k) = {\cal P}_{\zeta, sr}(k)
 \left( 1 + {\cal P}_{\zeta, sr}(k) f_2(\xi) e^{4\pi\xi} \right),
\end{equation}
\begin{equation}
{\cal P}_{\zeta, sr}(k) =
 \left( \frac{H^2}{2\pi \dot \varphi } \right) ^2.
\end{equation}
The function $f_2(\xi)$ is defined in \cite{Barnaby:2010vf, Barnaby:2011vw}.

\begin{figure}[!t]
\center %
\includegraphics[width=9 cm, trim = 0 0 0 0 ]{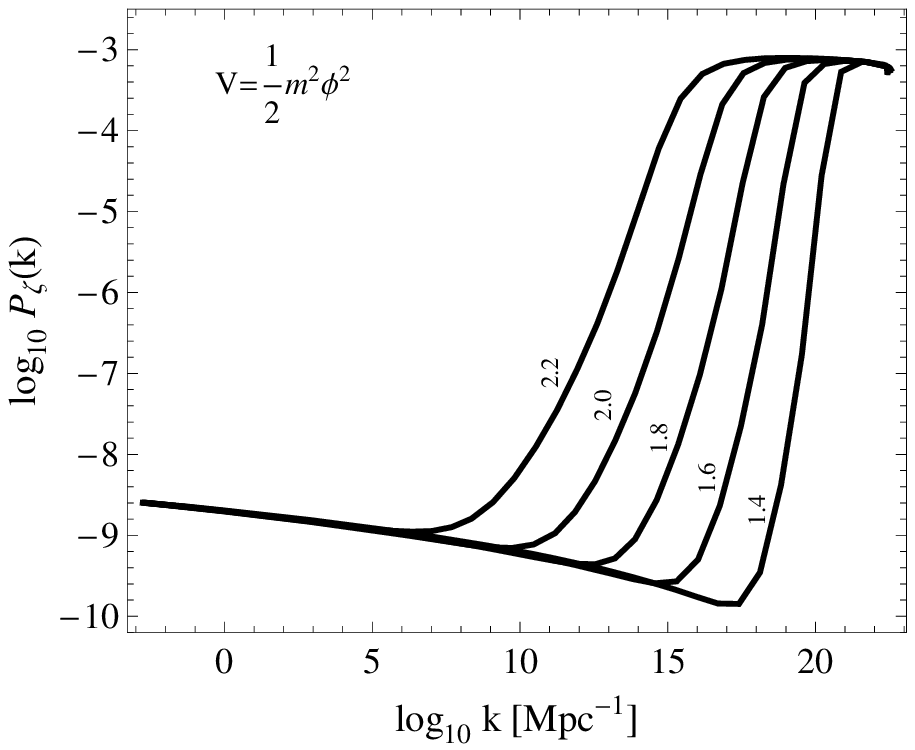}
\includegraphics[width=9 cm, trim = 0 0 0 -10 ]{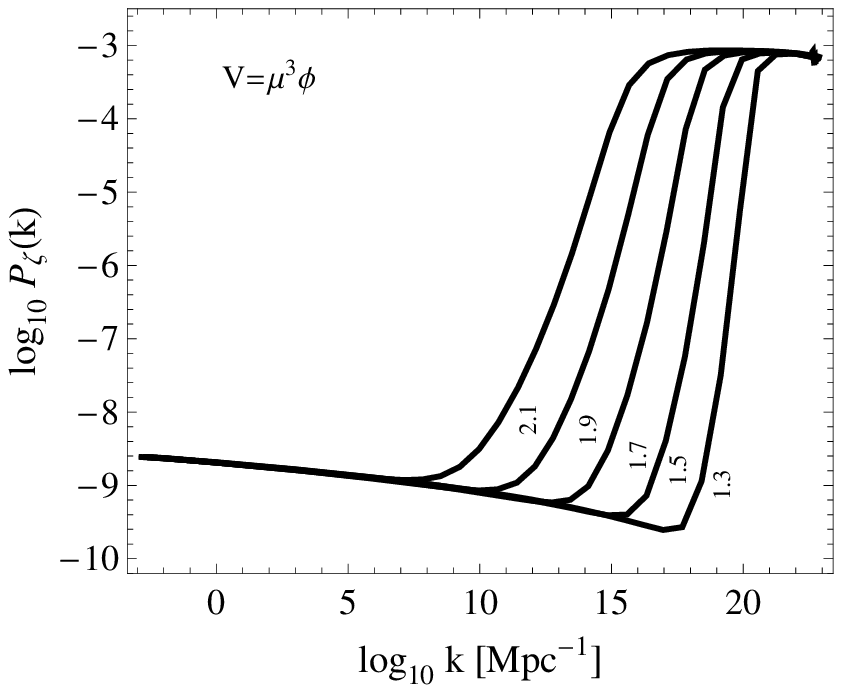}
\caption{ \label{fig-P}
The curvature perturbation power spectrum ${\cal P}_\zeta (k)$ calculated for
different values of $\xi_{CMB}$, for two shapes of inflaton
potential. The curves are labeled with the value of $\xi(N_{CMB})$.
}
\end{figure}

Near horizon crossing one has the approximate solution of Eq. (\ref{ddotbeta}) (everywhere
below we omit the contribution of the vacuum part, i.e., the solution of the
homogenous equation):
\begin{equation}
\delta \varphi \approx \frac{ \alpha }{f}
\frac{\left( \vec E \cdot \vec B - \langle \vec E \cdot \vec B \rangle \right) }
{3 \beta H^2 },
\end{equation}
and, correspondingly, one has for the curvature (see Appendix \ref{sec-app-hor}):
\begin{equation} \label{zetaappminusalf}
\zeta \approx - \frac{ \alpha }{f}
 \frac{  \left( \vec E \cdot \vec B - \langle \vec E \cdot \vec B \rangle \right) }
{3 \beta H \dot \varphi}.
\end{equation}
The variance of the curvature power spectrum is \cite{Linde:2012bt}
\begin{equation} \label{e20}
\langle \zeta(x)^2\rangle = \frac{H^2}{\dot\varphi^2} \langle \delta \varphi^2\rangle
 \approx \frac{\alpha^2}{f^2}
\frac{\langle \vec E \cdot \vec B \rangle ^2}{(3 \beta H \dot\varphi)^2}.
\end{equation}
From this equation, in the limit of large backreaction, when $\beta\gg 1$, the simple
approximate formula for the power spectrum is obtained
\cite{Anber:2009ua, Barnaby:2011qe, Linde:2012bt}:
\begin{equation}
{\cal P}_\zeta(k) \approx \langle \zeta(x)^2\rangle = \frac{1}{(2 \pi \xi)^2}.
\end{equation}
Some examples of the power spectrum solutions are shown in Fig. \ref{fig-P}.
For the calculations we used the approximate formula (\ref{e20})
which takes into account backreaction (at latest stages of inflation
the backreaction effects are quite essential).
Everywhere we add the contribution of the vacuum part which is
dominant at small values of $k$. The connection of the comoving wave number
$k$ with $N$ is given by
\begin{equation}
k=a_e H(N) e^{-N},
\end{equation}
where $a_e$ is the scale factor at the end of inflation.

\section{PDFs and non-Gaussianity}
\label{sec-3}

For a derivation of the PBH constraints we need an expression for the PDF
of the $\zeta$-field. Evidently, this is a technical problem in non-Gaussian case
because, for a calculation of the PDF one must know, in principle, all
cumulants (moments) contributing to its series expansion.

In our case, the simplest assumption which we can use in this concrete model is
the following \cite{Linde:2012bt}:
$\zeta$-field is distributed as a square of some Gaussian field $\chi$,
\begin{equation} \label{A3}
\zeta = \chi ^2 - \langle \chi ^2 \rangle,
\end{equation}
having in mind that non-Gaussianity of fluctuations $\delta\varphi$,
described by, e.g., Eq. (\ref{ddotbeta}), arises just from the fact
that the particular solution of this equation is bilinear in the field $A_\mu$
(the latter is assumed to be Gaussian).

If Eq. (\ref{A3}) holds (in this case, we have so-called $\chi^2$-model),
the PDF of $\zeta$ is given by (see, e.g., \cite{Bugaev:2013vba, Bugaev:2012ai})
\begin{equation}
\label{A7}
p_\zeta(\zeta) = \frac{1}{ \sqrt{\zeta + \langle \chi^2 \rangle} } \;
p_\chi\left(\sqrt{\zeta + \langle \chi^2 \rangle}\right) \; ,
\end{equation}
\begin{equation} \label{A4}
p_\chi(\chi) = \frac{1}{\sigma_\chi \sqrt{2\pi}} e^{-\frac{\chi^2}{2 \sigma_\chi^2}}, \qquad
\sigma_\chi^2 \equiv \langle \chi^2 \rangle.
\end{equation}
Variance and skewness of the $\zeta$-field are, respectively,
\begin{equation}
\langle \zeta^2\rangle = 2  \langle\chi^2\rangle^2, \qquad
\langle \zeta^3\rangle = 8  \langle\chi^2\rangle^3,
\end{equation}
so that the first non-trivial reduced cumulant is
\begin{equation} \label{D3}
D_3 = \frac {\langle \zeta^3\rangle } {\langle \zeta^2\rangle^{3/2}} = \sqrt{8} .
\end{equation}

More generally, one can use $\chi^2_n$-model, in which the $\zeta$-field is written as
a sum of $n$ squares of Gaussian fields,
\begin{equation} \label{chisqsum}
\zeta = \sum_{i=1}^{n} \chi_i ^2 - n \langle \chi_i ^2 \rangle.
\end{equation}
In this case, the PDF of $\zeta$ is \cite{Matsubara:1995ns, Koyama:1999fc}:
\begin{equation} \label{pnun}
p_\nu(\nu) = \frac{\left(1 + \nu \sqrt{\frac{2}{n}} \right)^{\frac{n}{2}-1}}
{\left(\frac{2}{n}\right)^{\frac{n-1}{2}} \Gamma\left(\frac{n}{2} \right)}
\exp\left( - \frac{n}{2} \left( 1 + \sqrt{\frac{2}{n}} \nu\right) \right),
\end{equation}
\begin{equation}
\nu \equiv \frac{\zeta}{\sqrt {\langle \zeta^2 \rangle} }, \qquad
p_\nu(\nu) d\nu = p_\zeta(\zeta) d\zeta.
\end{equation}
The cumulants of $\chi^2_n$-distribution are given by the simple formula,
\begin{equation} \label{D3new}
D_m = (m-1)! \left( \frac{2}{n}\right)^{\frac{m}{2}-1}.
\end{equation}
It is tempting to assume that the best choice in our case is the $\chi^2_2$-model,
i.e., $n=2$, in accordance with the fact that the photon has two polarizations.
The expression for the corresponding PDF follows from Eq. (\ref{pnun}):
\begin{equation} \label{pnu22}
p_\nu(\nu) = e^{-(1+\nu)},
\end{equation}
and the PDF for the $\zeta$-field is
\begin{equation}
p_\zeta(\zeta) = \frac{1}{\sqrt{\langle \zeta^2 \rangle} } p_\nu(\nu),
\end{equation}
with properties
\begin{equation}
\int\limits_{   -\sqrt{\langle \zeta^2 \rangle}    }^{\infty}
\zeta p_\zeta(\zeta) d\zeta = 0 ;  \qquad
\int\limits_{   -\sqrt{\langle \zeta^2 \rangle}    }^{\infty}
p_\zeta(\zeta) d\zeta = 1 ;  \qquad
\int\limits_{   -\sqrt{\langle \zeta^2 \rangle}    }^{\infty}
\zeta^2 p_\zeta(\zeta) d\zeta =  \langle \zeta^2 \rangle.
\end{equation}


If a PDF of the $\zeta$-field is known one can calculate not only the reduced
cumulants $D_m$ but also shapes of  $\zeta$-polyspectra (e.g., shapes of $\zeta$-bispectra).
From the other side, some of these functions can be calculated in our inflation
model directly, without using the PDF. In particular, the reduced cumulant $D_3$ is
given by the simple relation \cite{Linde:2012bt} (in the region where the backreaction
is large):
\begin{equation}
D_3 = \frac {\langle \zeta^3\rangle } {\langle \zeta^2\rangle^{3/2}} \cong
\frac{ {1 / (4\pi^3\xi^3)} } { {(1 / (2 \pi \xi)^2)^{3/2}} } = 2.
\end{equation}
This value coincides with the $D_3$ following from Eq. (\ref{D3new}) for $n=2$
(compare it with the $D_3$ value given by Eq. (\ref{D3})).
So, the choice of $\chi_2^2$-model for a description of PDF seems to
be appropriate.

Some results of $\zeta$-bispectrum calculations in our axion inflation model
and a comparison with corresponding $\chi^2$-model predictions are given
in the Appendix \ref{sec-app-bisp}.

\section{PBH constraints}
\label{sec-4}

For calculations of PBH constraints we need PDF for the {\it smoothed}
$\zeta$-field, $\zeta_R$ ($R$ is the smoothing radius). We assume,
using the argumentation of works \cite{Peebles:1998ph, White:1998da, Koyama:1999fc}
(see also \cite{Bugaev:2012ai}) that PDF of the smoothed $\zeta$-field can be
expressed in the form
\begin{equation}
p_{\zeta,R}(\zeta_R) = \frac{1}{\sqrt{\langle \zeta_R^2 \rangle }} \tilde p_{\tilde \nu}(\tilde \nu),
\quad
\tilde \nu = \frac{\zeta_R}{\sqrt{\langle \zeta_R^2 \rangle }}.
\end{equation}
Besides, we assume, following conclusions of \cite{Seto:2001mg, White:1998da}
that cumulants of PDF are approximately equal in smoothing and non-smoothing cases,
\begin{equation} \label{DDapp}
D_{m,R} \approx D_m.
\end{equation}
It follows from Eq. (\ref{DDapp}) that PDF of the smoothed $\zeta$-field can
be written as \cite{Bugaev:2012ai}
\begin{equation}
p_{\zeta,R}(\zeta_R) = \frac{1}{\sqrt{ \langle \zeta_R^2 \rangle }} p_{\tilde \nu}(\tilde \nu),
\end{equation}
where $p_{\tilde \nu}(\tilde \nu)$ is given by Eq. (\ref{pnun}) with $n=1$ for $\chi^2$-model
and $n=2$ for $\chi^2_2$-model, with a substitution $\nu \to \tilde \nu$.
In this approximation the effects of the smoothing come only through the variance
$\sqrt{ \langle \zeta_R^2 \rangle}$ while the shape of the PDF is the same as
in the non-smoothing case. The variance of $\zeta_R$ is given by the formula
\begin{equation}
\langle \zeta_R^2 \rangle = \int\limits_0^{\infty} \tilde W^2(kR) {\cal P}_\zeta (k) \frac{dk}{k},
\end{equation}
where $\tilde W (kR)$ is a Fourier transform of the window function \cite{LL2009},
and we use a Gaussian one, $\tilde W^2(kR) = e^{-k^2 R^2}$.

One can show that the energy density fraction of the Universe contained in PBHs
which form near the time of formation, $t=t_f$ (at this time the horizon mass is
equal to $M_h(t_f) = M_h^f$) is given by the integral \cite{Bugaev:2012ai, Bugaev:2011wy}
\begin{eqnarray} \label{qq1}
\label{omPBH-beta}
\Omega_{PBH}(M_h^f)
\approx \frac{1}{\rho_i} \left( \frac{M_h^f}{M_i} \right)^{1/2} \int n_{BH}(M_{BH}) M_{BH}^2 d \ln M_{BH}
\approx \nonumber  \\ \approx
\frac{(M_h^f)^{5/2}}{\rho_i M_i^{1/2}} n_{BH}( M_{BH}) \left. \right|_{ M_{BH} \approx f_h M_h^f} .
\end{eqnarray}
Here, $n_{BH}(M_{BH})$ is the PBH mass spectrum,
$\rho_i$ and $M_i$ are, correspondingly, the energy density and horizon mass
at the beginning of radiation era (if the reheating is fast, it coincides with an end of inflation).
$f_h$ is the constant [equal to $(1/3)^{1/2}$] which connects the value of PBH mass forming at the
moment $t_f$ with the horizon mass at that moment (see, e.g., \cite{Bugaev:2008gw}).
The PBH mass spectrum in Press-Schechter \cite{PS} formalism is proportional to the derivative
$\partial P/ \partial R$, where $P$ is the integral over the $\zeta$-PDF \cite{Bugaev:2011wy},
\begin{eqnarray} \label{PRzc}
P(R) = \int \limits_{\zeta_c}^\infty p_\zeta d \zeta.
\end{eqnarray}
Approximately, one has \cite{Bugaev:2012ai, Bugaev:2011wy}
\begin{eqnarray} \label{qq2}
\Omega_{PBH}(M_h^f) \approx \beta_{PBH}(M_h^f),
\end{eqnarray}
where $\beta_{PBH}$ is, by definition, the fraction
of the Universe's mass in PBHs at
their formation time,
\begin{eqnarray}
\beta_{PBH}(M_h^f) \equiv \frac{\rho_{PBH}(t_f)}{\rho(t_f)} .
\end{eqnarray}
Now, having Eqs. (\ref{qq1}, \ref{qq2}), one can
use the experimental limits on the value of
$\beta_{PBH}$ \cite{Kohri:1999ex, Carr:2009jm} to constrain parameters of models used for
PBH production predictions. The PBH mass spectrum needed for a derivation of
$\Omega_{PBH}$ in Eq. (\ref{omPBH-beta}) depends on the amplitude of the curvature power spectrum
${\cal P}_\zeta$ (see \cite{Bugaev:2011wy, Bugaev:2012ai, Bugaev:2013vba} for details).

\begin{figure}[!t]
\center %
\includegraphics[width=10 cm, trim = 0 0 0 0 ]{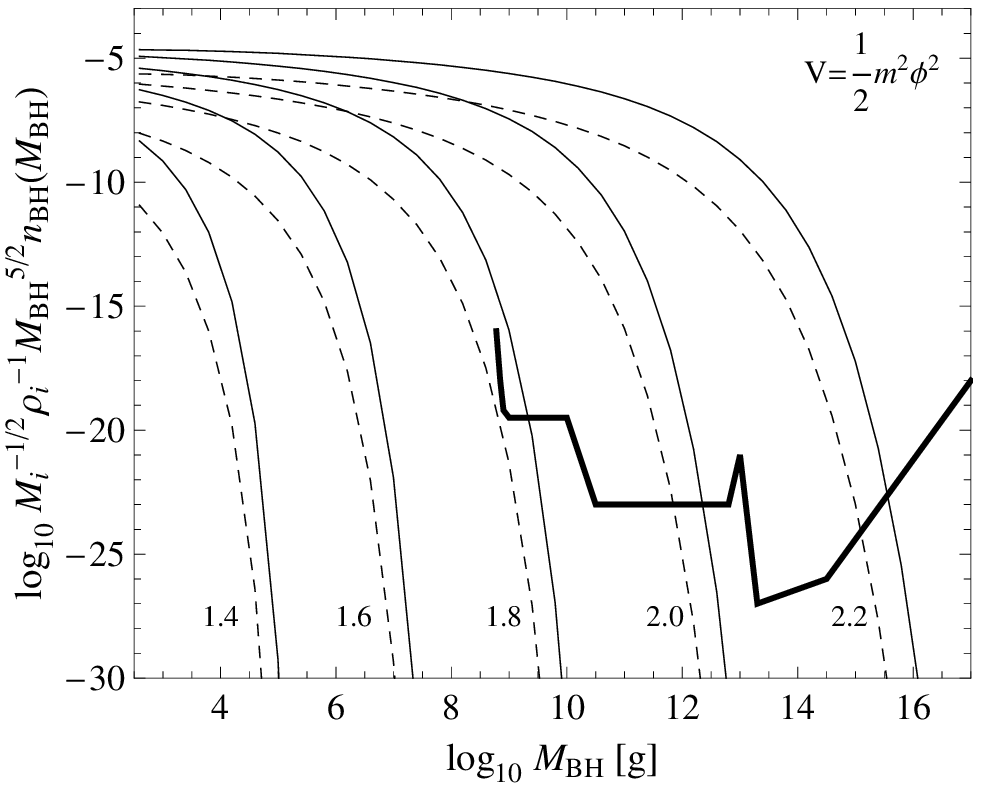}
\includegraphics[width=10 cm, trim = 0 0 0 -10 ]{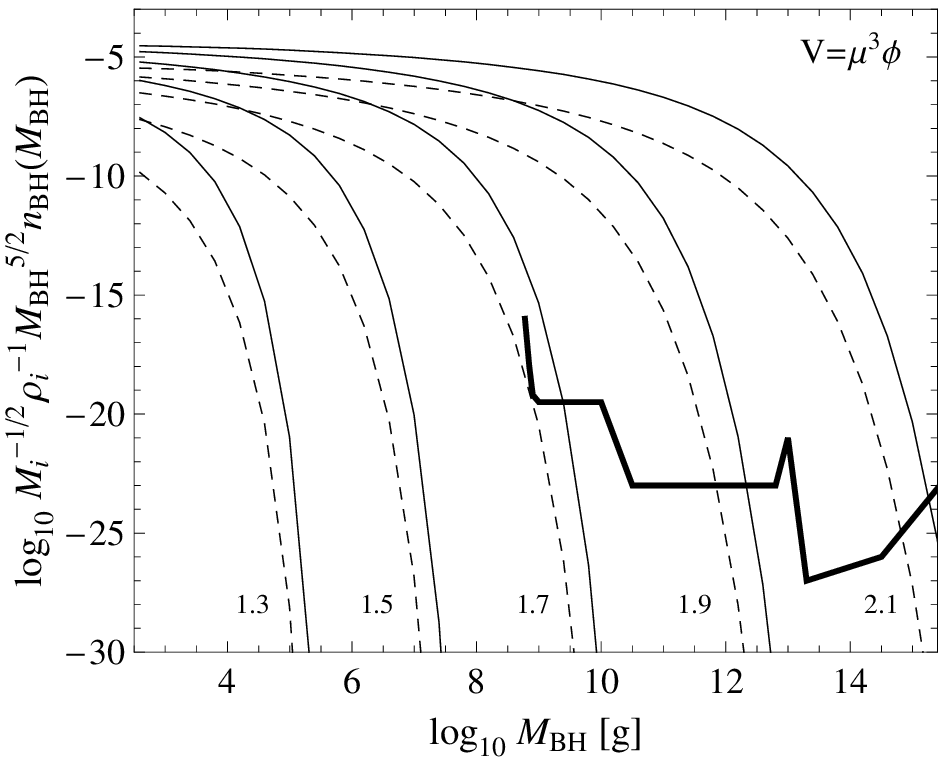}
\caption{ \label{fig-nBH}
Primordial black hole mass spectra corresponding to curvature perturbation power spectra
shown in Fig. \ref{fig-P}. Solid curves are for the case $\zeta_c=0.75$ while dashed
curves are for  $\zeta_c=1$. The curves are labeled with the value of $\xi(N_{CMB})$.
Thick line schematically shows existing constraints on PBH
abundance \cite{Kohri:1999ex, Carr:2009jm}.
}
\end{figure}

The results of PBH mass spectra calculation for the considered model are given in
Fig. \ref{fig-nBH} for several values of the parameter $\xi_{CMB}\equiv \xi(N_{CMB})$
and for two choices of the parameter $\zeta_c$, which is a model-dependent PBH
formation threshold (see, e.g., \cite{Bugaev:2011wy}).
For a calculation of the $\zeta$-PDF entering Eq. (\ref{PRzc}) we used $\chi^2_2$-model.

\begin{figure}[!t]
\center %
\includegraphics[width=8 cm, trim = 0 0 0 0 ]{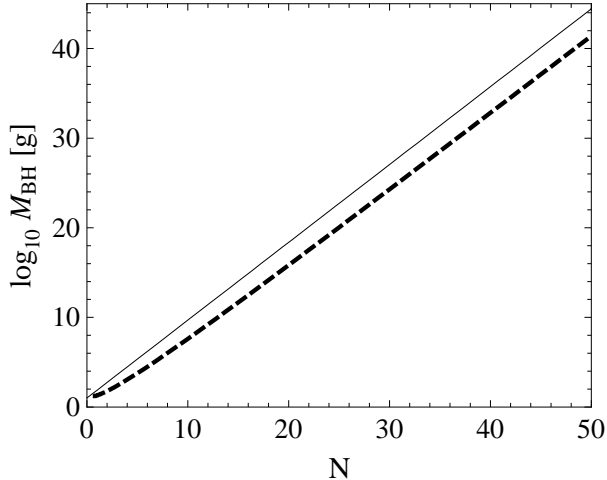}
\caption{ \label{fig-MBH}
Primordial black hole mass $M_{BH}$ that is produced, depending on the number $N$ of inflation e-folds.
Dashed line is calculation using our formulas (\ref{MBH-eq}), solid line is obtained using formula
$M_{BH}=10 e^{2 N}$g of work \cite{Linde:2012bt}.
}
\end{figure}

The PBH mass value, as a function of $N$, in our model is given by
the formula
\begin{eqnarray}
\label{MBH-eq}
M_{BH} = \frac{f_h M_{eq} k_{eq}^2}{a_e^2}  \frac{ e^{2 N} } { H(N)^2 } ,
\end{eqnarray}
where $H(N)$ is the Hubble constant during inflation at the epoch determined by
the value of $N$, $a_e$ is the scale factor at the end of inflation,
$M_{eq}$ and $k_{eq}$ are horizon mass and wave number corresponding to the moment
of matter-radiation equality.
The result of the calculation using Eq. (\ref{MBH-eq}) is shown in
Fig. \ref{fig-MBH} together with the result of the calculation using the more simple formula
suggested in \cite{Linde:2012bt} (namely, $M_{BH}=10 e^{2 N}$g). It is seen that
the curves start at almost the same value at $N=0$. The difference at larger
$N$ is due to the fact that Eq. (\ref{MBH-eq}) takes into account the
dependence of $H$ on $N$.

\section{Results and discussion}
\label{sec-results}

The main results of the paper are shown in Figs. \ref{fig-P} and \ref{fig-nBH}.
Fig. \ref{fig-P} illustrates the fact that due to tachyonic instability
of gauge field, an amplitude of the curvature power spectrum is very
large (up to $10^{-3}$) at small scales, $k \sim (10^{15} - 10^{20})$Mpc$^{-1}$,
for a broad range of $\xi_{CMB}$ values.
Fig. \ref{fig-nBH} shows the PBH mass spectra for definite values of the parameter
$\xi_{CMB}$.
On the vertical axis of Fig. \ref{fig-nBH} the combination
$M_i^{-1/2} \rho_i^{-1} M_{BH}^{5/2} n_{BH}(M_{BH})$ is shown; just this combination
is approximately equal to $\beta_{PBH}$, as it follows from Eq. (\ref{omPBH-beta}).
We compare these spectra with PBH data \cite{Kohri:1999ex, Carr:2009jm},
in which we consider only data for $M_{BH}> 10^9\;$g, as most reliable ones.
For such a comparison we drew in Fig. \ref{fig-nBH} the zigzag line representing,
schematically, the well known $\beta_{PBH}$-constraint summary
curve (see Fig. 9 in Ref. \cite{Carr:2009jm}). If some of our curves crosses
this zigzag line, the corresponding $\xi$-value is, according to our logic, forbidden.
Finally we obtain the constraint on the value of $\xi_{CMB}$, for
quadratic potential (\ref{V-square}),
\begin{eqnarray}
\xi_{CMB} < 1.8 \;. \label{lim-sq}
\end{eqnarray}
This constraint can be compared with the corresponding result of the
work \cite{Linde:2012bt}, $\xi_{CMB} < 1.5$.
In terms of $\alpha$ and $f$ constants, the limit (\ref{lim-sq}) corresponds to $\alpha/f < 26 M_P^{-1}$.

We performed similar analysis for the case of linear potential (\ref{V-linear}), and
in this case the constraint on $\xi_{CMB}$ turns out to be
more strong,
\begin{eqnarray}
\xi_{CMB} < 1.7 \;,
\end{eqnarray}
corresponding to $\alpha/f < 36 M_P^{-1}$.

For a derivation of these results, we used the assumption
that $\zeta$-field has a $\chi^2_2$-distribution. For a comparison, we also performed the
same calculations for a simple $\chi^2$-model (with one degree of freedom) and obtained
the following PBH limits on the parameters: for the quadratic potential
$\xi_{CMB} < 1.75$, and for the linear potential $\xi_{CMB} < 1.65$.
Luckily, the constraints weakly depend on a choice of PDF ($n=1$ or $n=2$).

One should note, in conclusion, that PBH constraints are stronger than those from
CMB scales \cite{Pajer:2013fsa} and forthcoming constraints from gravity
wave experiments \cite{Crowder:2012ik}.

\section*{Acknowledgments}
The work of P.K. was partially supported by the grant of President of RF number SS-3110.2014.2.


\appendix
\section{Curvature power spectrum behind the Hubble horizon}
\label{sec-app-hor}

It is well known that, in general, the curvature  perturbation amplitude
$\zeta$ doesn't stay constant in time after its scale exits the horizon during
inflation. It is so even in the standard single-field inflation model if, in
particular, slow-roll is temporarily violated in a process of the inflationary
expansion \cite{Seto:1999jc, Leach:2000yw, Leach:2001zf} . It had been shown in
 \cite{Leach:2000yw, Leach:2001zf} (see also \cite{Bugaev:2008bi}) that in such models
the modes can have a very complicate evolution and can be strongly amplified on
super-horizon scales. As a result of such amplification, in particular, the
perturbation amplitudes at horizon re-entry can differ rather strongly from amplitudes
at a time of the exit.

In this Appendix we derive the curvature perurbation power spectrum following,
closely, the work \cite{Anber:2009ua}. Two main differences are:
{\it i)} authors of \cite{Anber:2009ua} assume that $\alpha$ is very large
($\sim 10^2$ or larger), and {\it ii)} they considered a case of the cosine
potential [given by Eq. (\ref{Vphi4})]. In contrast with this, we considered
the case when $f/\alpha \ll M_P$, $\alpha\sim 1$ and our potentials are
nonperiodic. We show in this Appendix that, nevertheless, the resulting
spectrum formula in our case is just the same as in \cite{Anber:2009ua} if
we limit ourselves to a consideration of the small scales, exiting the horizon at
final stages of inflationary expansion. Just these scales are of interest for
us because we study the PBH production processes.

The Eq. (\ref{ddotbeta}) which takes into account the back-reaction effects
can be simplified using the slow roll approximation in the background
equation (\ref{ddotphi}). We assume that the slow roll regime is supported,
mainly, by the dissipation into gauge field modes, i.e.,
\begin{equation}
3 H \dot \varphi \ll V' \;,  \label{A1}
\end{equation}
\begin{equation} \label{A2}
V' \cong \frac{\alpha}{f} \langle \vec E \cdot \vec B \rangle \;.
\end{equation}
The inequality (\ref{A1}) holds if $f/\alpha$ is small compared with $M_P$.
Using the definition of $\xi$ [Eq. (\ref{xi-equiv})] and the approximate
relation $3 H^2 M_P^2 \cong V$ one can rewrite (\ref{A1}) in a form
\begin{equation}
2 \xi \cdot \frac{f}{\alpha} \cdot \frac{V}{V'} \ll 1 \;.   \label{A3a}
\end{equation}
For the quadratic potential, $V=\frac{1}{2}m^2\varphi^2$, one obtains from (\ref{A3a}):
\begin{equation}
\xi \cdot \frac{f}{\alpha} \varphi \ll 1 \;,   \label{A4a}
\end{equation}
and, for the linear potential, $V=\mu^3\varphi$,
\begin{equation}
2 \xi \cdot \frac{f}{\alpha} \varphi \ll 1 \;.   \label{A5}
\end{equation}

We are interested in the final stage of inflation when small scales exit the horizon
($N\sim 10$). During this stage $\xi \sim 5$ (see Fig. \ref{fig-xi}) and
$\varphi \sim M_P$ \cite{Linde:2012bt, Barnaby:2011qe}. Substituting in Eqs.
(\ref{A4a}) and (\ref{A5}) our limiting values of $f/\alpha$ (see Sec. \ref{sec-results})
one can see that inequalities (\ref{A4a}) and (\ref{A5}) really hold.

To obtain the approximate equation (\ref{A2}) one must show that the term $\ddot\varphi$
in Eq. (\ref{ddotphi}) is small in comparison with $V'$. The proof of it is easily performed
in complete analogy with the proof of (\ref{A1}). Now, using (\ref{A2}) and the relation
following from Eq. (\ref{beta-equiv}),
\begin{equation}
\beta = 1 - \pi \langle \vec E \cdot \vec B \rangle  \frac{\alpha^2}{3 H^2 f^2},  \label{A6}
\end{equation}
one can rewrite Eq. (\ref{ddotbeta}) in the form (changing the time variable on
$\tau \cong -1/(aH)$ and going over in a $k$-space) \cite{Anber:2009ua}
\begin{equation}
\delta\varphi''(\vec k) - \frac{2}{\tau}
\left( 1 + \frac{\pi \alpha V'}{2 f H^2} \right) \delta\varphi'(\vec k) +
\left( k^2 + \frac{V''}{H^2 \tau^2} \right) \delta\varphi(\vec k) =
-\frac{\alpha}{f} a^2 {\cal J}(\tau, \vec k) \; , \label{A7a}
\end{equation}
\begin{equation}
{\cal J}(\tau, \vec k) = \int \frac{d^3 x}{(2\pi)^{3/2}} e^{-i \vec k \vec x}
\left[ \vec E \cdot \vec B - \langle \vec E \cdot \vec B \rangle  \right] \; .
\end{equation}

We can treat $V'/H^2$ and $V''/H^2$ as adiabatically evolving parameters,
as well as $H$ and $\xi$
(e.g., for for the quadratic potential, $V'/H^2 \sim (V'/V)M_P^2 \sim M_P^2/\varphi$,
$V''/H^2 \sim M_P^2/\varphi^2$), because $\Delta \varphi \ll \varphi$ over
$\Delta t \sim H^{-1}$  \cite{Linde:2012bt, Barnaby:2011qe}.
Due to this, we neglect their time dependence during the essential part of the
inflationary evolution of each mode. In this case the homogenous equation
(\ref{A7a}) (i.e., the equation (\ref{A7a}) with ${\cal J}(\tau, \vec k) = 0$)
can be written in a form
\begin{equation}
\tau^2 \delta\varphi'' + b \tau \delta\varphi' + (c \tau^2 + d)\delta\varphi = 0,
\end{equation}
\begin{equation}
b = - \frac{\pi \alpha V'}{f H^2}-2 \; , \;\; c=k^2 \; , \;\; d=\frac{V''}{H^2}.
\end{equation}
The solution of this equation is expressed through the cylindrical functions
(see, e.g., \cite{Kamke}):
\begin{equation}
\delta\varphi = \tau^{\frac{1-b}{2}} Z_\nu(k\tau), \;\; \nu=\frac{1}{2}\sqrt{(1-b)^2 - 4d},
\end{equation}
\begin{equation}
Z_\nu(k\tau) = C_1 J_\nu(k\tau) + C_2 N_\nu (k\tau),
\end{equation}
\begin{equation}
N_\nu (k\tau) = \frac{J_\nu(k\tau) \cos(\pi\nu)  -  J_{-\nu}(k\tau)} {\sin (\pi\nu)}.
\end{equation}
One can check using estimates given above that $|b|\gg 1$, $d \ll |b|$, so
\begin{equation}
\nu \approx \frac{1}{2} (1-b) \sqrt{1 - \frac{4 d}{b^2} } \approx \frac{1-b}{2} - \frac{d}{1-b}.
\label{A13}
\end{equation}
We are interested in the power spectrum at $k \ll aH$, i.e., at $k|\tau| \ll 1$, so, one can use
the approximation
\begin{equation}
J_\nu(x) \approx \left( \frac{x}{2} \right)^\nu \frac{1}{\Gamma(\nu+1)}. \label{A14}
\end{equation}
The solution of the full equation (\ref{A7a}) is obtained by the
variation of constants method (or, that is technically the same,
by the method of Green functions) and is given by the integration
over the source
function ${\cal J}(\tau, \vec k)$. Using the approximation (\ref{A14}) one
obtains, finally,
\begin{equation}
\delta\varphi \sim -\frac{\alpha}{f} \int\limits_{-\infty}^{\tau} d\tau' \tau'
\left\{ \left( \frac{\tau}{\tau'}\right)^{\nu+\frac{1}{2}-\frac{b}{2}}  -  \left( \frac{\tau}{\tau'}\right)^{-\nu+\frac{1}{2}-\frac{b}{2}}
 \right\}
a^2(\tau') {\cal J}(\tau', \vec k).
\end{equation}
Since $|\tau|<|\tau'|$ one can neglect the first term in figure brackets,
because $\nu+\frac{1}{2}-\frac{b}{2} \approx 1-b \gg 1$,
$-\nu+\frac{1}{2}-\frac{b}{2} \approx \frac{d}{1-b} \ll 1$.
It leads, with
using (\ref{A13}), to
\begin{equation}
\delta\varphi \sim \frac{\alpha}{f} \int\limits_{-\infty}^{\tau} d\tau' \tau'
\left(\frac{\tau}{\tau'}\right)^{\frac{d}{|b|}} \;,
\end{equation}
\begin{equation}
\frac{d}{|b|}  \approx \frac{V'' f }{\pi \alpha V'} \ll 1.
\end{equation}
Using this expression and the relation $\zeta=H(\delta \varphi / \dot \varphi)$, a
formula for the curvature perturbation power spectrum is
obtained straightforwardly \cite{Anber:2009ua}, with the result:
\begin{equation}
{\cal P}_\zeta \approx \frac{10^{-2}}{\xi^2} \left( \frac{\xi k}{a H} \right)^{\frac{2 d}{|b|}},
\;\; k \ll aH. \label{A19}
\end{equation}
We see from this formula that the power spectrum at super-horizon scales  has
no amplification, on the contrary, it decreases with a time when the scale moves
away from the horizon. Due to a small value of $d/|b|$ the time dependence is rather
mild.
Further, we see from Eq. (\ref{A19}) that in a limit of small $d/|b|$ which
corresponds to a limit of  the large back-reaction, the curvature
spectrum is almost scale invariant in a region of small scales, in
accordance with the results shown in Fig. \ref{fig-P}.
We come to a conclusion that our estimates
of the spectrum amplitude based on the approximate solution of Eq. (\ref{ddotbeta})
[given by Eq. (\ref{zetaappminusalf})] are reliable.

\begin{figure}[!t]
\center %
\includegraphics[width=9 cm, trim = 0 0 0 0 ]{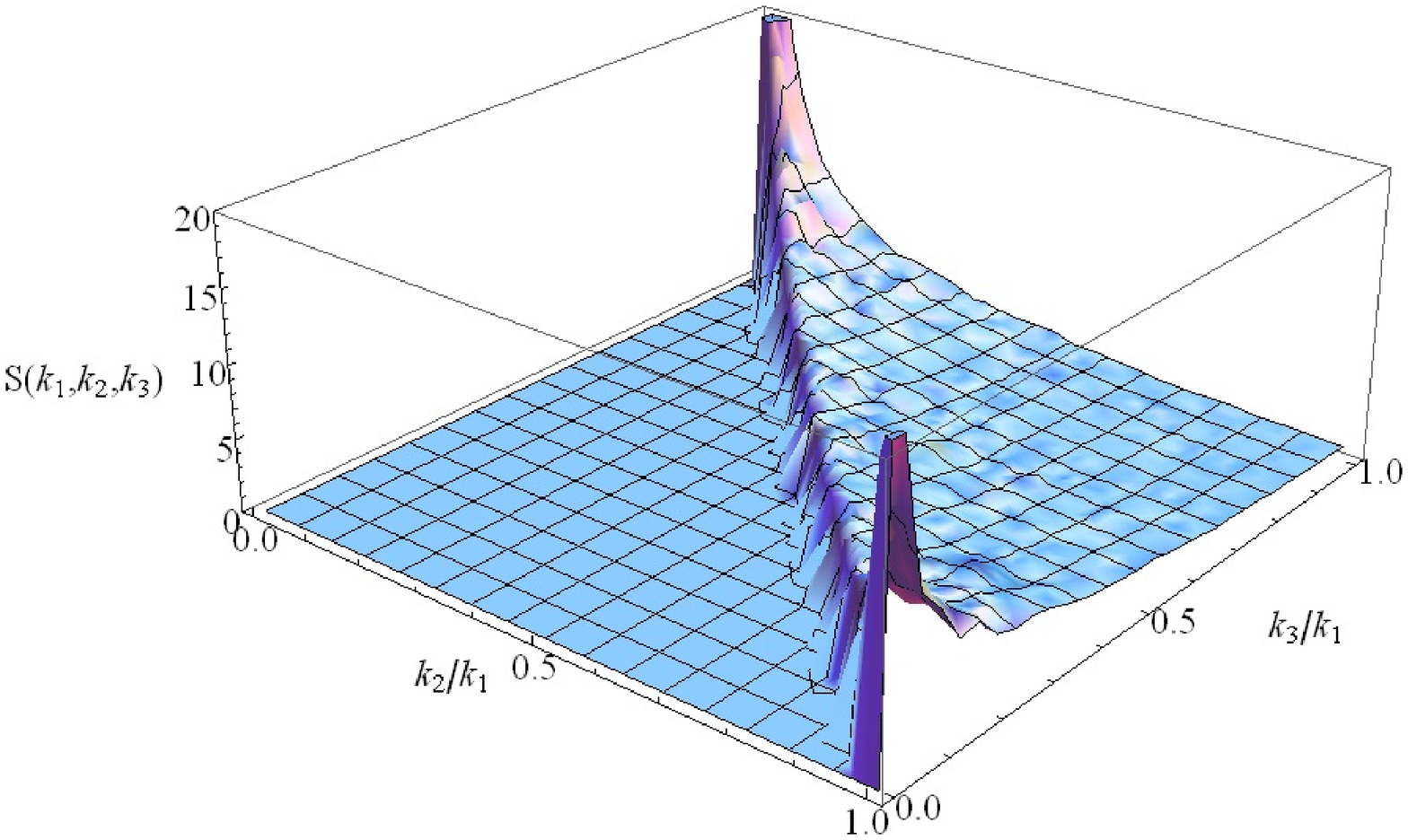}
\includegraphics[width=9 cm, trim = 0 0 0 -10 ]{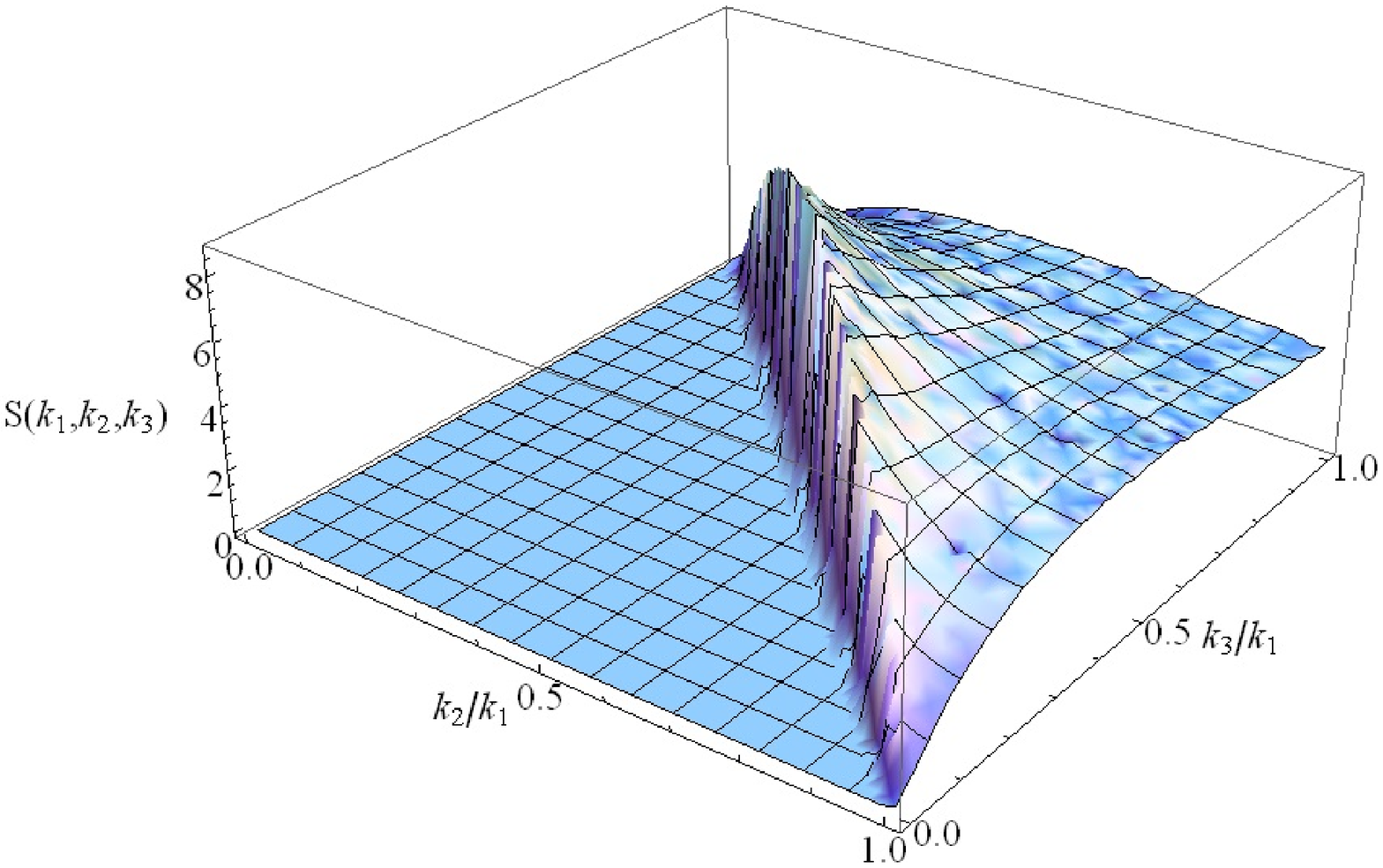}
\caption{ \label{fig-S}
Shape functions $S(k_1,k_2,k_3)$ (arbitrarily normalized) for $\chi^2$-model (upper panel) and
axion inflation model (lower panel).
}
\end{figure}

\section{The shape of the $\zeta$-bispectrum}
\label{sec-app-bisp}

The bispectrum of the non-Gaussian $\zeta$-field is defined by the expression
\begin{equation}
\langle \zeta({\bf k_1}) \zeta({\bf k_2}) \zeta({\bf k_3}) \rangle = (2\pi)^3
\delta({\bf k_1} + {\bf k_2} + {\bf k_3}) B({\bf k_1}, {\bf k_2}, {\bf k_3}).
\end{equation}
If $\zeta=\chi^2 - \langle\chi^2\rangle$, the formula for $B$ is \cite{LoVerde:2013xka}
\begin{equation}
B({\bf k_1}, {\bf k_2}, {\bf k_3}) = \frac{8}{3}\left[ \int \frac{d^3 k'}{(2\pi)^3} P_G(|{\bf k_1}-{\bf k'}|)
P_G(|{\bf k_2}+{\bf k'}|) P_G(k') + {\rm 2 \; perm.} \right],
\end{equation}
where $P_G(k)$ is the curvature power spectrum of the Gaussian $\chi$-field, $P_G(k)\sim k^n$.
The shape $S$ of the bispectrum, which is defined by the formula
\begin{equation}
S(k_1,k_2,k_3) = (k_1 k_2 k_3)^2 B(k_1,k_2,k_3)
\end{equation}
has a characteristic ``squeezed'' form, shown in Fig. \ref{fig-S} (upper panel; $n=-2.9$).

The bispectrum in our axion inflation model is calculated using the formula
\begin{equation}
B(k_1,k_2,k_3) = \frac{3}{10} {\cal P}_{\zeta, sr}^3 e^{6\pi \xi}
\frac{k_1^3+k_2^3+k_3^3}{k_1^3 k_2^3 k_3^3}
f_3\left(\xi, \frac{k_2}{k_1}, \frac{k_3}{k_1}\right).
\end{equation}
Here, the function $f_3$ is defined in \cite{Barnaby:2011vw, Meerburg:2012id}.
The example of the calculation of the corresponding shape function (for $\xi=6$)
is shown in Fig. \ref{fig-S} (lower panel).

Comparing two shape functions, one can see that the shape function of
our model differs rather strongly from the typical equilateral shape function
(see, e.g., \cite{Wang:2013zva} for examples of equilateral shapes).
At the same time, there is some similarity with the $\chi^2$-model
prediction (on both figures there is some concentration of points along the
diagonal line).

\end{document}